\input iopppt.tex
%
\pptstyle
\title{`Operational' Energy Conditions}

\author{Adam D Helfer}

\address{Department of Mathematics, Mathematical Sciences Building, 
University of Missouri, Columbia, Missouri 65211, U.S.A.}

\jl{6}

\beginabstract
I investigate the negative energy densities predicted by relativistic
quantum field theories.  Combining the quantum inequalities of Ford and
Roman, which restrict the time for which a negative energy density may
persist, with quantum limitations on measuring devices, I show that a
Klein-Gordon field in Minkowski space obeys the \it operational weak
energy condition:  \rm the energy of an isolated device constructed to
measure or trap the energy in a region, plus the energy it measures or
traps, cannot be negative.  There are good reasons for thinking a
parallel result holds locally for linear fields in curved space-times.
Indeed, there is reason to think an operational dominant energy
condition is satisfied.

\cabs 
A specific thought experiment to measure energy density is
analyzed in some detail, and the operational positivity is clearly
manifested.

\cabs
If such operational energy conditions are valid generally,
a case can be made that space-time cannot be adequately modeled
classically when negative energy density effects are significant.
\endabstract


\date

\def\gtrsim{\gap}
\def\lesssim{\lap}
\def\T{{\widehat T}}

\def\tint{{\int _{-\infty}^\infty{\widehat T}_{00}(t,0,0,0) b(t)\d t}}
\newcount\EEK
\EEK=0
\def\eek{\global\advance\EEK by 1\eqno(\the\EEK )}
\newcount\FOOT
\FOOT=0
\def\fnote#1{\advance\FOOT by 1\footnote{${}^{\the\FOOT}$}{#1}}
\def\jtemjtem#1{\par\indent #1\enspace\hangindent2\parindent
\ignorespaces}

\section{Introduction}
A startling prediction of relativistic quantum field theory is that,
while the total energy of a system should be positive or zero, the energy
density, and hence the energy of a subsystem, can be 
negative~(Epstein et al 1965).
And indeed this possibility is present generically.
Even for a Klein-Gordon field on Minkowski space, for any
smooth compactly supported bump function $B$, the expectation values
$$\langle\int {\widehat T}_{00}B \, \d ^3x\rangle\, ,\eek$$
for ${\widehat T}_{00}$ the renormalized
Hamiltonian density, are unbounded below~(Helfer 1996).  Thus
states with arbitrarily negative energy densities are always
available.  The set of states with this expectation value equal to
$-\infty$ is dense in the Hilbert space.

Negative total energy densities have never been directly observed.
Although the Casimir effect is sometimes cited as a case where a
negative energy density has been measured, in Casimir-type experiments
it it one of the `pressure' components of the stress-energy operator
(${\widehat T}_{zz}$) which is measured, and not the energy density
(${\widehat T}_{tt}=\T _{00}$).  There is an indirect mathematical relation
between them (the long-time average of the force can be computed as
minus the gradient of the energy), but they are different quantum
observables.  Just such distinctions will be central in this paper.
The emphasis will be on the question of what direct local measurements
of $\T _{tt}$ are possible.

Still, negative total energy densities have received extensive
theoretical investigation because they contravene a basic tenet of
classical physics.  Indeed, if there were no restrictions on the
negative energies achievable, there would be gross macroscopic
consequences:  an ordinary particle could absorb a negative energy and
become a tachyon; an isolated patch of negative energy would give rise
to a repulsive gravitational field; one could violate the second law of
thermodynamics by using negative energies to cool systems without an
increase of entropy~(Ford 1978, 1991; Davies 1982; Grove 1988); and the
general-relativistic effects might include traversable wormholes, `warp
drives' and time machines~(Morris and Thorne 1988, Morris et al 1988,
Alcubierre 1994, Everett 1996, Ford and Roman 1996, Pfenning and Ford
1997).  Too, it is something of a puzzle why such states do not
interfere with the dynamics of the quantum fields:  why do not
perturbations (which are always present) send the field cascading
through these negative-energy states, with a corresponding release of
positive-energy radiation?  It is a matter of common experience that
such effects do not occur, or at least not often, and therefore there
must be some mechanism restricting the production of negative energy
densities, their magnitudes, durations, or interactions with other
matter.

Such a mechanism was first proposed by Ford (1978), and has been
investigated by him and Roman~(Ford 1991, Ford and Roman 1997
and references therein).
They establish restrictions on the negative energy density and flux that
can persist for a time, and present arguments
that  (for free Bose fields in Minkowski space) the
negative energy $-\Delta E$ localizable in a time of order
$\Delta t$ should satisfy a \it quantum inequality \rm
$$\Delta E\Delta t\lesssim\hbar\, .\eek$$\xdef\frek{\the\EEK}%
These inequalities are powerful; they evidently limit the occurences of
negative energies considerably.  However, they do not as they stand seem
to be a full explanation.  For one thing, the argument for (\frek ) depends 
on a certain `coherence' assumption, which is
not generally valid.\fnote{Ford (1991)
recognized that (\frek ) could be violated if
the coherence condition does not hold.  A weaker
bound he
establishes (our inequality (4), below),
which is suggestive of the averaged weak energy condition in
general relativity, does not depend on this condition.
This weaker condition serves to severely limit wormholes and `warp
drives' (Ford and Roman 1996, Pfenning and Ford 1997).}
For another, it is not clear that simply
restricting the occurences of large negative energies to short times is
enough to rule out unphysical effects.  Indeed, explicit analyses of
attempts to violate the second law of thermodynamics indicate that while
the quantum inequalities play a key role, an equally important one is
played by limitations on the measuring devices.  (See Ford 1978, 1991;
Davies 1982; Grove 1988. 
These also point out that identifying the 
characteristic time $\Delta t$ which is relevant for a
particular physical problem may not be an easy matter.)

In Section 2 of this paper, I shall re-examine the derivation of the
quantum inequalities, find a way to repair the `coherence' condition,
and argue further that a device constructed to measure or capture a
local negative energy $-\Delta E$ must itself have energy at least \rm
$\Delta E$.  One may say briefly that \it operationally the total
energy must be non-negative.  \rm In Section 3, I shall extend the
reasoning to establish an operational dominant energy condition for
free fields in Minkowski space.  This says, roughly, that the total
four-momentum density, of the field plus the measuring device, must be
future-pointing.

In Section 4, I shall describe a thought-device which measures energy
density, and discuss its quantum limitations.  For this device, I find
even more stringent restrictions than the operational energy
conditions.  The Planck scale appears explicitly as a limitation on its
measurements, and below the Planck scale timing errors prevent one from
approaching the regime where the negative energy densities are
comparable to the energy density of the device.
(See Ford et al 1992, Ford and Roman 1993 for previous discussions of
detectors.)

In Section 5, I shall show that 
at least in two limiting regimes,
in a thought-experiment using this
device to measure the energy density between the plates of
a Casimir apparatus, the operational positivity of energy still holds.
(Other regimes would require better quantum inequalities on the Casimir
configuration than are presently known.)
This is remarkable, as the Ford--Roman inequalities do not apply on this
scale, and indeed treat negative energy densities which arise in a
distinct way from the negative Casimir energy.

Section 6 gives a more extensive discussion of what operational
positivity of energy means, assesses the likelihood that it holds
universally, and analyzes its implications for gravity.  One can make a
serious argument that space-time cannot be adequately
modeled by a classical manifold with classical metric in regimes where
negative energy densities are significant.

The final section summarizes the main conclusions.

The metric signature is ${+}\,{-}\,{-}\,{-}$.

\section{The Quantum Inequalities}
Ford and Roman have
given several derivations of the quantum inequalities,
but the elements which are relevant here are common to all.  Consider
the quantity
$${\cal E}=-\inf _{|\Phi\rangle}\langle\Phi |\int _{-\infty}^\infty
  {\widehat T}_{00}(t,0,0,0) \, b(t)\, \d t |\Phi\rangle\, ,\eek$$
(over normalized states $|\Phi\rangle$ in the Fock space of a
Klein-Gordon field in Minkowski space), where
$b(t)=t_0/[\pi (t^2+t_0^2)]$
is a `sampling function' with integral unity and characteristic width
$t_0$.\fnote{The
sampling function is peaked over an
interval of characteristic width $\sim t_0$, but not supported only there.  
It is not possible to localize to 
a sharply demarcated time interval, because the quantity
$\langle\Phi |{\widehat T}_{00}(t,0,0,0)|\Phi\rangle$ is a distribution, 
and turns out to involve terms like $\delta '(t)$.
It is this which is behind
difficulties in identifying the correct $\Delta t$ for physical
applications.}  It is shown that 
$${\cal E}\leq k\hbar c(ct_0)^{-4}\, ,\eek$$\xdef\Feek{\the\EEK}%
where it is known that the numerical
constant $k\leq 3/(32\pi ^2)$.  
Up to this point the
argument is essentially mathematical.

The next step is physical.  If a device were to be constructed to
measure or trap this negative energy within an interval of length $t_0$,
then in order to function coherently the linear dimension of the device
must be no larger than $ct_0$.\fnote{Perhaps
one should use $ct_0/2$ here.  This would make the force of our later
arguments somewhat stronger.}  Thus the magnitude of the negative
energy within the device
$$\Delta E\leq {\cal E}\cdot (4\pi /3)(ct_0)^3\leq (4\pi k/3)\hbar {t_0}^{-1}
  \leq \hbar /(8\pi t_0)\, .\eek$$\xdef\feek{\the\EEK}%
This hypothesis of coherent functioning deserves closer scrutiny.

The trouble here is that although it may be reasonable to think of an
experiment as a whole (including preparation at the start and collection
of data at the end) as `coherent' on a time scale $T_0$, the scales
$t_0$ of the components of the experiment may be much smaller.
For example,
suppose we had $N$ devices obeying (\feek ), so
capable of detecting or
trapping a negative energy
$-\varepsilon(4\pi k/3)\hbar {t_0}^{-1}$ 
in time $t_0$; here $\varepsilon <1$ is the
efficiency of the device.
These devices are arranged in
in an array in space, and in a common rest-frame.  Each carries a clock
which has been synchronized with (say) a master clock in the center of
the array.  At a preset time, each device operates.  Then, if the field
is in a suitable configuration, the total negative energy absorbed will be
$-N\varepsilon(4\pi k/3)\hbar {t_0}^{-1}$
and the interval will be $t_0$.  (Note that there is no
requirement that the devices be near one another, so they can be
separated far enough apart that locality considerations guarantee that
the quantum field can indeed be in a state which will produce such a
negative energy.  Also note that while it is true that construction of
the array of devices requires a different time scale than $t_0$, that
time scale is \it larger, \rm namely the time required
to synchronize the devices, greater than $\sim N^{1/3}t_0$.)  
By choosing $N$ large enough, we can arrange for an arbitrarily large
negative energy to be trapped within a time $t_0$.  
Thus even if we start from `coherent' devices, we can create others
which violate the quantum ineqality (\feek ).

We can repair this by taking into account the energy of the measuring
device.
A device which measures $\tint$ must involve some sort of clockwork
mechanism and transducer which function to weight 
the contributions of ${\widehat T}_{ab}$ at different times by $b(t)$. 
This clockwork must be able to resolve time increments of order $t_0$.
(Actually, in order to treat the function $b(t)$ as free from quantum
indeterminacy, the temporal resolution must be finer.)  Now let us
recall that a clock mechanism which is accurate to a time of order $\Delta
t$ must have mass $\gtrsim \hbar /(c^2 \Delta t)$ and so energy
$E_{\rm mech}\gtrsim\hbar /\Delta t$ (Salecker and Wigner 1958).  
For any one clock having energy $E_{\rm mech}$, then, controlling a
measuring device, the inequality (\feek ) applies with $t_0\sim\hbar
/E_{\rm mech}$.

This suggests that any device controlled by a clock of energy
$E_{\rm mech}$ can detect or trap negative energies $-\Delta E$ with
$-\Delta E+E_{\rm mech}\geq 0$ only.  
It should be made clear that this argument is not a mathematical proof.
For one thing, the
quantities $\Delta t$ and $E_{\rm mech}$ are only defined as orders of
magnitude, and it is in this sense that $E_{\rm mech}\gtrsim\hbar /
\Delta t$ is known to hold.  
For another, the quantum inequality (\feek ) as only been established
for one form of sampling function.  Nevertheless, the numerical factor
$1/8\pi$ in inequality (\feek ) is far enough below unity that
it strongly suggests $E_{\rm mech}\geq\Delta E$.
For the remainder of this paper, I shall assume this is the case.

With this assumption, notice that a collection of
measuring or trapping devices deployed
and set to function simultaneously (or, more generally, at spacelike
separations), as in the example above, will also have total energy in
excess of the negative energy it can detect or trap.

One may summarize this contention by saying that \it operationally, the
energy must be non-negative, \rm that is, the sum of the measured
energy and the energy of the measuring device must be non-negative.

\section{The Dominant Energy Condition}
The treatment so far concerns the energy of a finite system, as measured
by an inertial observer.  
The result localizes:  even if one tries to separate the clockwork used for
measuring $\tint$ from the world-line $(t,0,0,0)$, one must still transmit
timing signals to the vicinity of this world-line, and these signals must
resolve times $\lesssim t_0$, which means the quanta carrying the signals must
have energies $\gtrsim \hbar /t_0$.  Thus locally the total energy, of
the field plus the measuring device, must be non-negative.
One may call this the \it operational weak energy
condition: \rm  $T_{ab}^{\rm op}t^at^b\geq 0$ for all timelike vectors $t^a$.

It is possible to derive a stronger result, the \it operational dominant
energy condition:  \rm  $T_{ab}^{\rm op}t^au^b\geq  0$ for all
future-pointing vectors $t^a$ and $u^a$.  The changes needed to the treatment
above are as follows.

\subsection{A Quantum Inequality for Momentum Density}
Let 
$$\Pi _a=\langle \int
_{-\infty}^\infty {\widehat T}_{a0}(t,0,0,0)b(t)\d t\rangle\, .\eek$$  
Then one can prove
$\Pi _a u^a\geq -(3/32\pi ^2)t_a u^a\hbar c/(ct_0)^4$ 
for any future-pointing vector
$u^a$.  Let
$$P_a=\langle \int
_{-\infty}^\infty \d t \int _{\| x\|\leq ct_0} \d ^3x\, 
  {\widehat T}_{a0}(t,x)b(t)\rangle\eek$$ 
be the expectation of the four-momentum measured in an experiment
controlled by a clock resolving times $\sim t_0$.\fnote{Inclusion of a
spatial bump function in the integral defining $P_a$ is possible, and
does not affect the argument.}  Then
$P_au^a\geq -(1/8\pi )t_au^a\hbar /t_0$.  Equivalently,
$$P_a= -(1/8\pi )(\hbar /t_0)t_a+\hbox{  a future-pointing vector}\,
.\eek$$\xdef\MEEK{\the\EEK}%

Only a few modifications to the analysis of Ford and Roman (1997) 
are needed to
establish this, and I shall simply indicate those, in the notation of
that paper.
The main point is that the Lemma of Appendix B in that
reference
remains valid when $p_j^*p_{j'}$ is replaced by any Hermitian matrix
$P_{jj'}$ which has non-negative eigenvalues, since each such matrix is
a sum of terms of the form $p_j^*p_{j'}$.  It holds in
particular when $j$, $j'$ are replaced by the wave-vectors ${\bf k}$,
${\bf k'}$ and
$P_{\bf k k'}=\bigl( 2u_{(a}t_{b)}k^a{k'}^b-u_at^ak_b{k'}^b\bigr)
/\sqrt{\omega\omega '}$.  The rest of the changes are self-evident.

\subsection{Constraints on the Measuring Device}
Consider a clock which may be boosted relative to $t^a$.  If the clock is
required to have resolution $\Delta t$ in the $t^a$-frame, then its
resolution in its own frame must be $\Delta t/\gamma$, with $\gamma$ the
usual Lorentz factor.  Its mass must be 
$$m\gtrsim\hbar\gamma /(c^2\Delta t)\, .\eek$$\xdef\mineq{\the\EEK}%
Let the clock's four-momentum $P_a^{\rm clock}$ be $(E,p)$ in the
$t^a$-frame, so $E=mc^2\gamma$ and $p=mc\beta\gamma$.  Then
$$E^2-mc^2E/\gamma -p^2c^2=0\eek$$
from which
$$\bigl( E-mc^2/(2\gamma)\bigr) ^2 -p^2c^2=m^2c^4/(4\gamma ^2)\, .\eek$$
This means
$$P^{\rm clock}_a=mc^2/(2\gamma )t_a +\pi _a\, ,\eek$$
where $mc^2/(2\gamma )\gtrsim \hbar /(2\Delta t)$ and $\pi _a$ is timelike
future-pointing with $\pi _a\pi ^a
=(mc^2/2\gamma )^2\gtrsim (\hbar /2\Delta t)^2$.

\subsection{The Operational Dominant Energy Condition}
Combining the results of the two previous subsections, we see that for any
future-pointing vector $u^a$, the sum of the expectation value $P_au^a$ of
the $u^a$-component of the momentum and the corresponding component of the
momentum of the clock which controls the sampling satisfies
$$\bigl( P_a+P^{\rm clock}_a\bigr) u^a 
    \geq mc^2/\gamma -(1/8\pi )\hbar /t_0\, ,\eek$$
which we expect to be positive by (\mineq ).

A word about the interpretation of this is in order.  Here $P_a$ is
the expectation of ${\widehat T}_{ab}t^b$ smeared over a volume in
Minkowski space.  The components of this smeared operator
do not generally
commute (one cannot simultaneously measure the components of the
four-momentum in a finite box, because of edge effects).  Thus it
perhaps too strong to say that the four-momentum is operationally
future-pointing, since the four-momentum of the field within a finite
box cannot, strictly speaking, be measured.  What I have shown is that
for any future-pointing vector $u^a$, the operator $u^aP_a^{\rm op}$ 
is non-negative,
where $P_a^{\rm op}$ is the sum, of the clock's four-momentum and the
four-momentum operator for the field in a box.

\section{A Model Measurement of Energy Density}
I shall describe here a thought-experiment to measure energy density
(and thus test the weak energy condition), and examine its quantum
limitations.  Of course, investigation of any one device cannot prove
that there are parallel limitations for all devices; but it does
provide a challenge to do better.

The measurement must depend on some coupling to the quantum field, and
in order to keep the model as realistic as possible and not make ad-hoc
assumptions about the coupling, I shall consider a device which detects
the energy density gravitationally. 

Suppose there are several nearby world-lines.  One is occupied by an
observer, who carries a clock, a photon source, and a photon detector.
The
other world-lines are those of mirrors.  The observer sends out pulses
of light to the mirrors, and measures the times it takes them to return. 
Each such measurement of time provides an estimate of the distance to
the mirror.  After two measurments (of light bouncing from the same
mirror), the observer can estimate a component of the relative velocity
of the mirror, and after three measurments she can estimate a relative
acceleration and hence one component of the curvature, $R_{0i0i}$ for a
mirror in the ${\imath}^{\rm th}$ direction.  By averaging over the
mirrors in the three spatial directions, she gets an estimate of
$R_{00}$.  Assuming the stress-energy is trace-free, this measures the
energy density.

(Ideally, one would like an experiment which directly measured the
energy density, without appealing to the trace-freeness of the
stress-energy.  Such devices can in principle be constructed by
considering four, relatively boosted, apparatuses of the sort described
above, and appropriately summing their outputs; similarly one could
test the dominant energy condition.  A device like this would be
subject to requirements at least as strict as those to be considered
here, unless one could find some cancellation of errors when summing
the outputs.  This would be rather complicated and will not be
considered further.  In any event, the trace of the stress-energy is
a c-number for electromagnetism which is second-order in the
gravitational field.)

A number of factors constrain the design of the device, and
limit its accuracy.  Let the accuracy of the clock be $\Delta t$, and
the time between bounces to the $\imath ^{\rm th}$ mirror be
$\sim t_i\geq\Delta t$.  Then:

\item{(a)} The clock's mass satisfies $mc^2\gtrsim (\hbar /\Delta t) (t/\Delta
t)$, where $t$ is its running time, $t\sim\max\{ t_1,t_2,t_3\}$
(Salecker and Wigner 1958).

\item{(b)} If the 
extents of the photons'
wave packets are not to contribute appreciably to timing errors,
the photons' energies must be $\gtrsim\hbar /\Delta t$.

\item{(c)} If the mirrors are not to be significantly accelerated by the
photons, each of their rest-energies must be $\gtrsim \hbar /\Delta t$.
(One could try to use less massive mirrors and correct
kinematically for the effects of the photons' impacts, but then
the photons are red-shifted after impact and there is a consequent
loss of accuracy.)

\item{(d)} The mirror and the observers must be outside each others'
Scwarzschild radii.  This means $ct_i\gtrsim Gm/c^2$.

\item{(e)} To measure the directions of the mirrors relative to the observer (in
order to properly weight the sum $\sum _{ij} R_{0i0j}g^{ij}$) 
to a given accuracy
$\Delta \theta$ requires an angular momentum $\gtrsim\hbar/\Delta
\theta$.

\medskip

One can see how these constraints enforce the operational positivity of
energy.  The apparatus measures the average energy density over a time
$\sim t$ and spatial extent $\sim ct$; by the Ford--Roman inequalities,
the magnitude of the negative energy density in a field is $|E_{\rm
neg}|\lesssim (1/8\pi )\hbar /t$.  However, the energy of the
apparatus must be greater than this, by (a), (b) and (c).

The full force of the constraints has not yet been used, and indeed they
imply more stringent limitations on the total energy density.

First, combining (a) and (d), one finds
$$\Delta t\gtrsim \bigl( G\hbar /c^5\bigr)^{1/2}=t_{\rm Planck}\, .\eek$$
Thus the Planck scale explicitly limits the
measurements.
Now we shall show that timing errors keep one from coming close to
violating positivity of energy, unless the Planck scale is approached.

The uncertainty in the measured curvature due to timing errors is
$\Delta R\sim \Delta t/(c^2t^3)$, leading to an uncertainty in the
measured energy of $\Delta E\sim \Delta R c^7t^3/G\sim c^5\Delta t/G$. 
Then
$$\Delta E/|E _{\rm neg}|\gtrsim c^5t\Delta t/(G\hbar)
    \gtrsim (c^5/G\hbar )(\Delta t)^2=(\Delta t/t_{\rm Planck})^2\, .\eek$$
This places severe restrictions on the possibility of observing total
energies which are `close to' negative.

\section{The Casimir Effect}
I shall now show that, at least in two limiting regimes, if the device
described above were used to measure the Casimir energy density, the
operational positivity of energy would still hold.
(Investigations beyond these regimes would require better quantum
inequalities for the Casimir configuration than are presently known.)
This is remarkable because:  (i) the Ford--Roman inequalities do not
hold in this case; and (ii) the negative Casimir energy density arises
in quite a different way from those considered in the Ford--Roman
analysis.

When a free field is quantized in a restricted volume with suitable
boundary conditions, the stress-energy operator takes the form
$${\widehat T}_{ab}={\widehat t}_{ab} +t^{\rm Casimir}_{ab}\; ,\eek$$
where ${\widehat t}_{ab}$ is a normal-ordered operator quadratic in the
fields, and $t^{\rm Casimir}_{ab}$ is a c-number, the Casimir-type
contribution.  The negative-energy density effects from the two terms
are distinct.\fnote{For a link between the two,
see the comments on the Casimir effect and quantum electrodynamics in
Section 6.2.}
Those from ${\widehat t}_{ab}$ are ultraviolet effects
which appear when the Hamiltonians in question do not correspond to
perfect symmetries of the theory.  (For example, measuring the energy in
a region corresponds to evolving only the field within the region.)  
But negative Casimir energies
represent a displacement of the vacuum relative to that of Minkowski
space.  For the original Casimir effect, two plane parallel perfect
conductors separated by a distance $l$ are found to have
$$t^{\rm Casimir}_{ab}=(\hbar c\pi ^2/720 l^4)
  \left[\matrix{-1&&&\cr &1&&&\cr &&1&\cr &&&-3\cr}\right]
  \eek$$\xdef\caseq{\the\EEK}%
between the plates,
in a Cartesian coordinate system adapted to the symmetry of the problem.

With this preamble, we can distinguish the two limiting regimes.

The first is the local one.  By this one means a measurement of the
average energy density over a space-time volume of spatial extent $\sim
a$ and temporal extent $\sim t$ such that $a,ct\ll l$ and the distance
of the volume to either plate is $\gg a,ct$.  One expects measurements in
such regions to be adequately modeled by those of free fields in
Minkowski space, 
and therefore the operational positivity to hold.

The second is the case of long times, that is, average energy density
measurements over times $\gg l/c$ (and over spatial volumes of some
finite size).  For sufficiently long times, one expects  fluctuations in
the field to average out, so the energy density can be well-modeled by
the c-number term.  (There will be spatial fluctuations, too, but over
long enough times one expects these to be negligible.  See Barton
1991a,b for an illuminating discussion.)  

Start by considering the effect of the clock's mass on the mirror in the
$z$-direction.  The curvature generated by the clock will be $\sim
Gm/(c^5t_z^3)$, whereas the Casimir energy density contributes a
curvature $\sim (G/c^4)\hbar c\pi ^2/720l^4$.  
Here $t_z$ is the time between photon bounces in the transverse
direction.
Since $ct_z\lesssim l$
and $mc\gtrsim\hbar /l$ (for the clock's Compton wavelength to be
smaller than $l$), the field due to the clock will dominate the
negative Casimir-energy effects.  This distortion is not averaged out
by summing over the spatial directions, and must be subtracted
to get a measurement of the energy density to the required accuracy.

To do the subtraction, one must know $mc^2/(ct_z)^3$ to an accuracy of
better than $\hbar c\pi ^2 /720 l^4$, that is, 
$\Delta \bigl( mc^2/(ct_z)^3\bigr)
\lesssim\hbar c\pi ^2/720 l^4$.  In particular this means $mc^3(\Delta
t_z)/(ct_z)^4\lesssim \hbar c\pi ^2 /720 l^4$, from which
$$mc^2\Delta t_z\lesssim \hbar (ct_z/l)^4\pi ^2 /720\lesssim \hbar \pi ^2
  /720\, .\eek$$
However, this would be a contradiction of the fundamental inequality of
Salecker and Wigner.

In both the local and long-time-average regimes, then, one expects
operational positivity of energy to hold for the configuration which
gives rise to the classical Casimir effect.

\section{Discussion}*
\subsection{What Operational Positivity Means}
Operational positivity of energy does not contradict the prediction of
negative energy-density states by relativistic quantum field theories.
For example, in the Casimir effect, there is no dispute that the energy
density operator is \it mathematically \rm not positive:  it has a good
mathematical definition, and its spectrum includes negative values.
Equally, there is no dispute that to compute the energy of the entire
Casimir system, of the plates plus the field, the negative binding
energy must be accounted for.  More generally, there is no dispute that
negative energy densities may contribute to lower the energy
of an unequivocally observable positive-energy object.
What operational positivity asserts is
that one could never 
\it locally experimentally \rm verify a \it total \rm
negative energy in a region.  Put another way, it asserts that certain
negative energy-density effects are shrouded in quantum measurement
problems.

It is important to keep in mind that it is something of an
abbreviation to speak of measurement of energy or energy density in the
present context.  What is really measured is the integral of $\T _{ab}$
against a sampling function.
The results of such measurements are not additive,
and so the sense in which an energy density
exists is less direct than in a classical theory.  In a given
experiment, one measures an \it average \rm of the stress-energy over a
given scale (set by the sampling function), 
and the same state, measured at different scales, would
generally give rise to different values of the energy density.  The
quantum inequalities lead us to expect that a state might have very
negative energy densities when measured on a sufficiently fine scale,
but only slightly negative energy densities on gross scales (cf.
Kuo and Ford 1993).

While the operational positivity condition is a limitation on
measurement, it is different from the usual limitations on
conjugate observables in three ways:

\item{(a)} It arises not from a failure of commutativity, but from the
time-energy uncertainty relation which requires a clock with a given
resolution to have a correspondingly great mass.  (This should be
distinguished from the relation which, for example, relates the widths of
spectral lines to the stability of the states.)  While this relation is
little discussed in texts (perhaps because it has no neat expression in
the usual mathematical formalism of quantum theory, where time is
only a parameter), its essence goes back at least to the
Bohr-Einstein dialog of 1930 (Bohr 1958).

\item{(b)} The inequality refers not just to the observables within the
quantum field theory, but also to the measuring device outside the theory.  

\item{(c)} The inequality cannot be expressed as a purely abstract
statement about operators on Hilbert space, but refers to the \it local
measurement \rm of the stress-energy:  it occurs because a clock, or
other timing signal, must be present \it where \rm the stress-energy is
measured.

To emphasize (c), there are cases in which
there is a clear sense in which a non-local measurement violates
operational positivity.  Consider an experimenter who prepares a
Casimir apparatus.  It is common in quantum theory to regard the
preparation of a state as a measurement (followed by a selection,
rejecting those states which are unsuitable).  Thus the preparation of
the Casimir experiment leaves the system in a state which is (to the
required accuracy) perfectly well known.  This state is an eigenstate of
the operator measuring the energy of the field between the plates, and
that energy is negative.  However, the measurements done in preparing
the state (adjusting the plates,
measuring their positions and verifying that they are held
steady while the field inside equilibrates) are \it not \rm local to the
space-time region between the plates after the field has equilibrated. 

It is an open question whether there are restrictions beyond the purely
local in the measurement of energy density.  For example, could one
infer from the gravitational multipole moments of an object, as
measured outside of it, that there was a negative energy density
somewhere inside?

At present, the only uncontroversial conclusions to be drawn from this
investigation are negative.  Any arguments depending on the
detection or absorption of negative energies should be re-examined to
see if they are affected by quantum measurement limitations.

However, when a limitation as a matter of
principle is discovered, there is the possibility that the physics of
the situation is not understood at a deep level.  In 
section~6.3, I shall argue that this is indeed the case, and that
space-time cannot be adequately modeled classically when negative
total energy density effects are significant.

\subsection{Generality of the Results}
The operational dominant energy condition has only been established for free
fields in Minkowski space.  (The investigation of the Casimir effect
concerned only the weak energy condition, and
concluded only that the specific device in question could not violate
positivity of energy --- it is possible that some other measurement can
be constructed which for which there is a violation.)
Nevertheless, the results are suggestive enough
that I

\bf Conjecture:  \rm
The operational dominant-energy condition is universally valid:
for any local measurement of the energy-momentum density one has
$T_{ab}^{\rm op}t^au^b\geq 0$, where $T_{ab}^{\rm op}$ is the sum of the
stress-energy of the field and of the measuring device.

Such a statement cannot be mathematically proved within the present
framework of quantum field theory.  The remainder of this subsection
assesses the evidence for and against the conjecture.

For linear fields on Minkowski space, one would expect to be able to
establish Ford-type inequalities, and the operational energy conditions
would follow.  For non-linear field theories, one would not expect such
simple, universal, Ford-type relations.  Still, the apparent validity
of operational positivity for the Casimir effect is circumstantial
evidence that the operational weak energy condition holds for quantum
electrodynamics.  The argument is this.  Presumably, the correct
first-principles treatment of the Casimir effect starts with a quantum
electrodynamic Lagrangian, including all of the ionic and electronic
structure of the conductors.  The usual, practical, calculation that is
done should be the result of integrating out the dynamically frozen
degrees of freedom from this first-principles treatment, and the
quantity $t_{ab}^{\rm Casimir}$ is the contribution of these frozen
degrees of freedom.  Thus the operational positivity in the Casimir
effect would be a special case of operational positivity in quantum
electrodynamics, and the validity of the former is circumstantial
evidence for the latter.

The most serious objection to the conjecture is that it could be
violated by simply having a sufficiently large number of elementary
field species.  However, the number required (if they
are all free fields) is $\gtrsim 8\pi$ (from
inequality (\feek )), and it is quite plausible that this is not achieved.

The operational weak or dominant energy conditions would hold for
linear fields in curved space-time if one had a curved-space analog of
the energy density inequalities (\Feek ) or (\MEEK ).  At present, no
general such result
is known, but it is generally believed that such results should exist.
It may be too much to hope for very general results with non-compactly
supported sampling functions, but one may reasonably

\bf Conjecture:  \rm
Let $b(t)$ be a smooth, compactly supported bump function of area unity.
Then, for a linear quantum field on a globally hyperbolic space-time,
one has
$$\langle\int _{-\infty}^\infty {\widehat T}_{ab}(\gamma (t))
  {\dot\gamma}^au^b \, b(t/t_0) \d t /t_0\rangle
    \geq -C(b){\dot\gamma}_a u^a t_0^{-4}\eek$$
for small enough $t_0$,
where $C(b)$ is positive and depends only on $b$, the curve $\gamma (t)$
is a timelike geodesic parameterized by arc-length increasing toward
the future and $u_a$ is a
future-pointing timelike vector, parallel-propagated along $\gamma (t)$.

While this has not at present been proved even in Minkowski space,
there are good if somewhat technical reasons for believing it, and for
expecting that $b$ can be chosen so that $C(b)$ is small (cf.~Flanagan
1997, Song 1997).  Then the arguments of the previous sections apply,
and the operational dominant energy condition holds in curved
space-time.

\subsection{Implications for Gravity}
Key to the development of both relativity and quantum theory was what
is now called operationalism:  the idea that the theory should be
formulated in terms of observables.  This criterion has been an
invaluable guide in the development of quantum field theory in curved
space-time, and should not be ignored.  I shall argue here that if we
adhere to it, and grant the operational weak energy condition, then
space-time cannot be adequately modeled classically in a regime where
negative energy densities are important.  While this argument is not
conclusive, it is strong enough to take seriously.

That negative total energy densities of themselves preclude a classical
model of space-time is a radical suggestion, since it implies that the
limit of validity of classical general relativity may be reached far
above the Planck scale.  I shall try to make clear exactly the sense in
which this limitation is supposed to occur.

Throughout, one must
remember that the energy density does not exist as a single operator,
but as an operator-valued distribution.  Thus the `regime' in
question refers not just to a region of space-time, but also to a
scale of measurement set by the sampling function $f^{ab}$ against
which $\T _{ab}$ is integrated.  That negative energy densities
are important in this regime means that the the projections of the
state onto the eigenspaces of $\int \T _{ab} f^{ab}\, \d {\rm vol}$
have significant components with negative eigenvalues.  The magnitudes
of these components may depend strongly on the scale over which $f^{ab}$
varies.  In an experiment, the function $f^{ab}$ is determined by the
apparatus used.

According to operationalism, we should ascribe a direct classical
meaning to a regime if there is, in principle at least, a means of
determining its structure by measurements, and that structure is
classical.  However, the operational weak energy condition, together
with Einstein's field equation, precludes any local, direct,
measurement of the geometry in a regime of negative energy density
(since the stress-energy can be inferred from the geometry).  In order
to understand the nature of the limitations involved in more detail, I
shall consider attempts to measure the geometry of space-time which are
on the verge of revealing negative energy densities.

I begin by
giving more precise statements of when a classical space-time may be
considered an adequate model.  This may be done by recasting Einstein's
arguments for a four-manifold structure a little.  I will distinguish
several different candidate definitions, in decreasing order of
strength:

\item{} We say that a space-time region is modeled by a classical
manifold with classical metric to a desired accuracy if:

\jtemjtem{(Definition a)} It is possible in principle to introduce test
particles (i.e., particles not interfering with the measurement) whose
trajectories can be measured to determine the geometry of the region
(to the desired accuracy).

\jtemjtem{(Definition b)} It is possible in principle to introduce
particles whose trajectories can be measured to determine the geometry
of the region.  The particles need not be test particles, but there
must be an unambiguous mathematical means of recovering the geometry of
the region (to the desired accuracy).

\jtemjtem{(Definition c)} It is possible in principle to consider a
sequence of similarly-prepared experiments in which particles are
introduced, their trajectories are measured and these statistically
determine the geometry of the region.  The particles need not be test
particles, but there must be an unambiguous mathematical means of
recovering, from the data for the entire sequence of experiments, the
geometry of the region in each experiment (to the desired accuracy).

\it Remark:  \rm In these definitions, the term `particle' does not
imply a classical point mass, but only an object with some degree of
localizability.  (One could contemplate definitions based not on
observations of elements of the particles' trajectories, but of other
quantum observables.  Such definitions, however, will not be considered
here.  They are less directly connected to the space-time geometry,
and so a proponent of such a definition would have to argue for an
additional interpretational liberty.)

One may interpret (a) as saying that a classical space-time has locally
verifiable properties which are independent of the precise means of
measurement.  In (b), the classical space-time exists, but attempts to
measure its geometry, while successful, in general yield results which
are dependent on the means of measurement:  one must introduce some
non-negligible amount of matter in order to effect the measurent,
and so the geometry of the region does not have an operational existence
independent of the means of measurement.
In (c), a sort of statistical element may be present as well.  (The
motivation for (c) will appear presently.)

Now let us examine what happens in a negative energy regime, that is,
when the accuracy of measurment is such that, on that length scale,
negative energy densities are significant.

Condition (a) is the strongest and arguably the closest to the usual
notion of a classical space-time.  Cases where it fails must be
considered to at least raise serious questions about the application of
classical general relativity.  And this condition is directly forbidden
by the operational energy conditions in a negative energy regime, since
the energy density is related to the geometry by Einstein's equation,
and a measurement of the geometry accurate enough to reveal a negative
energy density would require test particles and measuring apparatus of
a greater positive energy.  In essence, operational positivity
precludes the existence of \it test \rm particles in a negative energy
density regime, and this is why it is incompatible with (a).

Could space-time be classical in sense (b) or (c) when negative energy
densities are significant?  In each of these cases, the probing
particles must be about as massive as the source term whose geometrical
effect they seek to reveal.  Thus the problem of deducing the geometry
is one of strong-field gravity.  Recovering the space-time from the
data involves at least implicitly solving the joint evolution equations
for the field, the measuring device and the space-time.  This would be
difficult to pose properly and solve even if it were a classical
problem.  In the present context, though, the quantum nature of the
detector, because of its finite mass and hence finite Compton length,
as well as the quantum field observed, must be considered.

However, let us suppose a super-mathematician could overcome at least
the mathematical difficulties involved.  Still, there is good reason to
think that space-time will not be classical in the sense (b).
This is because the state of the measuring
device is subject to an uncertainty, manifested in the temporal
uncertainty $\Delta t$ in its resolving time.  It is uncertain precisely
when the apparatus will turn on and turn off, and this contributes to an
uncertainty in the measurement of the geometry.  This 
sort of timing error was analyzed at the end of Section~4, where it was
found that these dominate the measurement process unless one approaches
the Planck scale.  Of course, the analysis of Section~4 concerns only one
specific meauring device, and so there is the possibility that some
other sort of device might be able to overcome this difficulty. 

If one is willing to settle for a classical space-time in the sense
(c), then one can overcome the difficulty in the switching-on and
-off of the device by statistical means, since the structure of the
device and of the preparation of the experiment determines a
probability distribution of on- and off-times.  In other words, one
could similarly prepare an ensemble of space-times, quantum fields and
detectors, and examine the results of the experiments statistically.
Even so, it is not clear that the resulting data can be well-modeled
by a classical space-time or family of space-times.
This is because, in the present context, the different possible states
to which the state vector might reduce have differing geometries on a
scale which is significant for the problem.  (Even if the quantum
field's state does not have this property, the state of the detector
must, on account of its finite mass.)  Thus one is confronted with the
problem of understanding the dynamics of the reduction of the state
vector.  Whatever the solution of this is, there is no doubt of the
quantum character of the process.

The arguments above cannot be considered as rigorously proving that
space-time cannot be classical in negative energy density regimes.
But they reveal a number of quantum measurement problems.  How
serious these problems are is a matter of judgment.  Einstein created a
theory of ineffable beauty in General Relativity, and one should not
seek to move beyond this without the most scrupulous consideration.  But
one should have the same reservation about abandoning operationalism.

These comments apply in particular to the `semi-classical
approximation,' where the quantum
field and the space-time together satisfy the equation
$$G_{ab}=-8\pi G\bigl( T^{\rm classical\ matter}_{ab} +\langle 
  {\widehat T}_{ab}\rangle\bigr)\, .\eek$$\xdef\smceq{\the\EEK}%
The space-time manifold and its metric are treated classically, and the
quantity
$T^{\rm eff}_{ab}=T^{\rm classical\ matter}_{ab} +\langle 
{\widehat T}_{ab}\rangle$ acts as an effective classical source.
Kuo and Ford (1993) argued that negative energy-density states ought to
be characterized by large fluctuations, and this made it doubtful 
that equation (\smceq ) could be an approximation to a deeper
quantum theory of gravity and other fields.  The present analysis
uncovers another difficulty.
Only between measurements of the energy density (or when the effective stress
tensor satisfies the dominant energy condition), it is possible to
regard equation (\smceq ) (together with the
equation for the quantum field) as defining the evolution of the
space-time metric, including negative energy-density effects.  In
other words, \it
when negative energy-density effects are important,
the semi-classical equation can be
valid, as a local equation, only as long as 
there is no one there to check it! \rm

\subsection{An Example:  Cosmic Censorship}
The foregoing discussion dealt with generalities about the structure
of space-time.
I will close with a sketch of how
these considerations might apply to a specific issue, Cosmic Censorship.
As the Cosmic Censorship Conjecture is not settled at the
classical level, there is no pretense of my giving a complete
treatment of the matter.  However, even partial results are of interest,
and the argument given here exhibits the physics of the operational weak
energy condition briefly and clearly.

The Cosmic Censorship Conjecture asserts that, under physically
reasonable circumstances, space-time singularities are not visible from
great distances.  A commonly (but far from universally) held opinion on
this is that it ought to be true for `classical' matter fields (in
generic circumstances), but is unlikely to hold when the negative
energy densities engendered by quantum fields are taken into account.

Suppose that a singularity does arise in a region of negative energy
densities due to quantum fields.  We ask whether signals (say, photons)
revealing this singularity might escape to great distances. Now, while
there might well be \it mathematical \rm null geodesics escaping from
the singular region to infinity, the operational weak energy condition
would prevent them from carrying \it physical \rm photons energetic
enough to \it directly \rm reveal the geometry of the singular region.
(For then the photons' energy would have made the energy density in the
singular region positive.)  Thus a certain gross violation of the
Conjecture would be excluded.

Of course, this does not preclude the existence of other means of
detecting the singularity, and does not touch on the `classical'
question of whether positive energy-density singularities might be
visible from infinity, nor on whether negative energy densities in one
region could give rise to a singularity in another, positive energy
density, region.  The point is, whether Cosmic Censorship turns out to
hold in general or not, the physics of the scenario discussed in the
previous paragraph cannot be understood without considering quantum
measurement issues.

In the terminology of the previous subsection, this example shows the
inadequacy of a classical model in sense (a).  One might not be bold
enough to assert baldly that classical space-time has broken down, but
one must acknowledge the inadequacy of modeling the physics of
the situation simply by the geometry of the null geodesics.

\section{Conclusions}
I have suggested that quantum fields obey a new sort of energy
condition.  The \it operational weak energy condition \rm asserts that
locally, the energy of a field, together with the energy of an isolated
device measuring or trapping the field's energy, must be non-negative.  A
similar \it operational dominant energy condition \rm asserts that 
any timelike component of the total energy density must
be future-pointing.

There are good although at present not compelling arguments that at
least linear quantum fields obey an operational weak energy condition,
and perhaps an operational dominant energy condition.  The gaps in the
arguments come in part from technical difficulties in establishing
Ford--type quantum inequalities in the most general circumstances, and
partly from the order-of-magnitude nature of the inequalities of
Salecker, Wigner et al. on the masses of measuring devices.  Too, it
seems that the operational energy conditions could be violated if there
were too many elementary field species.

The operational dominant energy condition immediately resolves two of
the negative-energy pathologies listed in the introduction.  It
precludes the conversion of ordinary particles to tachyons.  And at the
level of Newtonian gravity, it forces gravitational fields to be
attractive in the sense that ${\bf\nabla}\cdot{\bf g}\leq 0 $, for
${\bf g}$ the gravitational acceleration field, since a measurement of
${\bf\nabla}\cdot {\bf g}$ is a measurement of the energy density.  The
remaining issues require more extensive discussion than can be given
here.  I only comment briefly that Grove~(1988), in his resolution
of the second-law problems raised by Ford~(1978) and Davies~(1982) in
effect establishes a special case of the operational positivity of
energy.  I hope to discuss this, and the question of why perturbations
do not cause quantum systems to decay into states with patches of
negative energy together with positive-energy radiation, in a future
publication.

My results cannot be considered good news for the
semi-classical approximation for gravity where there are
negative energy densities.  Even before this, Kuo and Ford (1993) had
argued that such states would have large fluctuations.  This means that
the expectations of the stress-energy do not reflect its eigenvalues,
and on general grounds one would not expect a semi-classical
approximation to be accurate.  Still, one might have thought the
approximation gives a rough classical response of the metric to the
quantum field.  But the present results restrict even this
interpretation:  where negative energy densities are significant, the
semi-classical approximation can be valid only as long as no one is
there to check it.  While validity in this sense
would be logically consistent,
it is at odds with the hard-headed operational world-view which
established the foundations of relativity and quantum mechanics.

The present results suggest that any attempt to understand the
consequences of negative energy densities for gravity (Hawking
evaporation; effect on singularity theorems, area theorem, positivity
of Bondi and ADM energies) must take into account quantum measurement
issues.

\ack
I am grateful to Larry Ford, Bahram Mashhoon, Tom Roman and the reviewers 
for useful comments.

\references

\refjl{Alcubierre, M 1994}{\CQG}{11}{L73}

\refjl{Barton, G 1991a}{\JPA}{24}{991}

\refjl{\dash 1991b}{\JPA}{24}{5533}

\refbk{Bohr, N 1958}{Atomic physics and human knowledge}{New York:  John
Wiley and Sons}

\refjl{Davies, P C W 1982}{\PL}{113B}{215-218}

\refjl{Epstein, H, Glaser, V and Jaffe, A 1965}{\NC}{36}{1016--22}

\refjl{Everett, A 1996}{\PR\ \rm D}{53}{7365}

\refbk{Flanagan, E 1997}{Quantum inequalities in two dimensional
Minkowski spacetime}{(gr-qc 9706006)}

\refjl{Flanagan, E and Wald, R M 1996}{\PR\ \rm D}{54}{6233}

\refjl{Ford, L H 1978}{\PRS}{A364}{227}

\refjl{\dash 1991}{\PR\ \rm D}{43}{3972-3978}

\refjl{Ford, L H, Grove, G and Ottewill, A 1992}{\PR\ \rm D}{46}{4566}

\refjl{Ford, L H and Roman, T A 1993}{\PR\ \rm D}{48}{776}

\refjl{\dash 1996}{\PR\ \rm D}{53}{5496}

\refjl{\dash 1997}{\PR\ \rm D}{55}{2082}

\refjl{Grove, P G 1988}{\CQG}{5}{1381-1391}

\refjl{Helfer, A D 1996}{\CQG}{13}{L129-L124}

\refjl{Kuo, Chung--I and Ford, L H 1993}{\PR\ \rm D}{47}{4510}


\refjl{Morris, M and Thorne, K 1988}{Am. J. Phys.}{56}{395}

\refjl{Morris, M, Thorne, K and Yurtsever, Y 1988}{\PRL}{61}{1446}

\refbk{Pfenning, M J and Ford, L H 1997}{The unphysical nature of
`warp drive'}{(gr-qc/9702026)}

\refjl{Salecker, H and Wigner, E. P 1958}{\PR}{109}{571-577}

\refbk{Song, Dae--Yup 1997}{Rectrictions on negative energ density in a
curved spacetime}{(gr-qc 9704001)}

\bye